# Elasto-capillary interaction of particles on the surfaces of ultra-soft gels: a novel route to study self-assembly and soft lubrication

Aditi Chakrabarti* and Manoj K. Chaudhury
Department of Chemical Engineering
Lehigh University, Bethlehem, PA 18015

**ABSTRACT.** We study the interaction of small hydrophobic particles on the surface of an ultra-soft elastic gel, in which a small amount of elasticity of the medium balances the weights of the particles. The excess energy of the surface of the deformed gel causes them to attract as is the case with the generic capillary interactions of particles on a liquid surface. The variation of the gravitational potential energies of the particles resulting from their descents in the gel coupled with the superposition principle of Nicolson allow a fair estimation of the distance dependent attractive energy of the particles. This energy follows a modified Bessel function of the second kind with a characteristic elastocapillary decay length that decreases with the elasticity of the medium. An interesting finding of this study is that the particles on the gel move towards each other as if the system possesses a negative diffusivity that is inversely proportional to friction. This study illustrates how the capillary interaction of particles is modified by the elasticity of the medium, which is expected to have important implications in the surface force driven self-assembly of particles. In particular, this study points out that the range and the strength of the capillary interaction can be tuned in by appropriate choices of the elasticity of the support and the interfacial tension of the surrounding medium. Manipulation of the particle interactions is exemplified in such fascinating mimicry of the biological processes as the tubulation, phagocytic engulfment and in the assembly of particles that can be used to study nucleation and clustering phenomena in well controlled settings.

*Email: adc312@lehigh.edu

## 1. INTRODUCTION

Capillary forces prevailing at the surfaces of liquids are exploited in various fields[1-27], a well-known example of which is the self-assembly of small particles into well-defined two dimensional structures[7-27]. Based on the studies spanning over several decades, the mechanisms underlying such interactions can be broadly classified into two major categories. If the bond number ($\Delta\rho g R^2/\gamma$, $\Delta\rho$ being the buoyant density of the particle, $\gamma$ is the surface tension of the liquid, $g$ is the acceleration due to gravity and $R$ is particle's radius) of a particle is significant, it can deform the surface of a liquid. In that case, the effective weight of a particle may be balanced by the capillary force of the deformed liquid surface if the particle is suitably hydrophobic. In delineating such interactions, Nicolson[7] showed that a superposition principle can be used by virtue of which the field energy is expressed in terms of the force that balances the effective weight of a single particle and the profile of the liquid surface that is deformed by the other proximous particle. The net result is that the energy of interaction follows[9-11] a modified Bessel function of second kind (zeroth order), the argument of which is the ratio of the distance of separation between the two particles and the capillary length of the liquid $L_c$ ($=\sqrt{\gamma/\rho g}$), $\rho$ being the density of the liquid. The Capillary length is an important material length scale in this problem that defines the range over which interaction prevails. Following the lead of Nicolson, several authors[9-11] derived detailed analytical expressions for the attraction between particles of spherical as well as cylindrical geometries on a liquid surface, which resolved some previous incomplete observations and analysis of Gifford and Scriven[8].

If the Bond numbers are very small, the particles can still interact[12,13,19] with each other on a liquid surface if the three phase contact line is uneven either due to chemical heterogeneity, rugosity or other anisotropies. In both cases of low and large particle bond numbers, however, the driving force is derived from the excess energy of the liquid surface intervening the particles. Kralchevsky and Nagayama[13] provided a detailed synopsis of the existing theories of capillary attractions and extended them to the interactions between particles embedded in thin films supported on a solid support. Significant progress has also been made in recent years that is based on the ideas of field theory[28] within the formalism of the Hilbert-Einstein action in a pseudo-Riemannian geometry. There are also various discussions in the literature[20-25] involving capillary interactions in conjunction with hydrodynamic and electrostatic forces leading to the self-assembly of colloidal particles.

Another class of interaction between particles arises due to the elastic forces of the surrounding medium, be it a thin membrane[29-31], an elastic string[32,33], a liquid crystal[34-37] or via an Eshelby type inclusions[38] in an elastic medium. In all these cases, the energetics of elastic distortion provides the necessary force for interaction. Interactions in liquid crystals are of recent interests, where a plethora of studies[35-37] following the lead of Poulin et al[34] shows that the distortion of the director fields in a liquid crystal around the dispersed particles give rise to a short range repulsion but a long range attraction. The orientational elastic energy dependent interaction in a liquid crystal also give rise to an interesting phenomenon as levitation[35,36] of nanoparticles above the ground surface. Elastic interactions in conjunction with defects in liquid crystals[37] have also been used to assemble particles in a pre-determined fashion. Combined effects of elasticity and capillarity have been observed in many two part systems, one of which provides the capillarity and the other elasticity[39-44]. However, to the best of our knowledge the literature is essentially devoid of the studies of interactions in a single medium that display both the above properties.

We recently reported[45,46] some experimental results related to the attraction of two spheres as they plunge deeply into a very soft hydrogel. Here as a sphere penetrates the gel, the neck of contact thins rapidly and forms a line singularity owing to the combined effects of a hoop stress and the adhesion forces thereby ensuring that the gel completely folds and self-adheres over the particle. The excess energy of the deformed surface coupled with the accumulated elastic strain energy is released as the particles approach each other, thus guaranteeing a net attraction[46] between the particles. This emergent[47] attractive force between the particles follows an inverse linear distance law that is long range and induces 3d assembly of small particles inside the gel. In this paper, we study the interactions of small particles that do not deeply plunge inside the gel; rather they float on its surface. This is a simpler system to consider for further exposition of the roles played by the elastocapillarity in the attraction of particles, which is also advantageous in studying the 2d self-assembly processes in a systematic way.

There are several motivating factors for such studies. In recent years[39-46,48-51], the roles played by elasticity, surface tension and the underlying hydrodynamics have captured the imaginations of physicists and engineers to study various phenomena of soft condensed matter. For example, one might be interested to understand how slender objects bend, fold, interact[40] and self-assemble with each other on the surface of a soft support (e.g. water). One may also be interested to develop a system to study the elasticity of particle rafts[48], kinetics of self-assembly of particles that form clusters or even jammed phases[49]. There are also several biological phenomena such as tubulation and phagocytosis that one may wish to mimic on the surface of a soft support[50,51]. Another interesting problem is that of soft lubrication[52,53], in which one is interested how frictional resistance to the motion of a particle develops in a system where elastic as well as hydrodynamic stresses act co-operatively. One envisages that many of the above mentioned phenomena would involve mesoscale level objects, which are amenable to detailed studies using an ordinary microscope and analysis using the models of continuum mechanics. Ordinary liquids with which capillary interactions are studied may not permit many of the mesoscale level objects to float on its surface if they are sufficiently denser than water. Furthermore, even if they float, there is very little control over the range, the strength as well as the time scales of such interactions. What we demonstrate in this work is that an ultrasoft hydrogel with a small amount of elasticity affords enough resistance to prevent sinking of a hydrophobic particle significantly denser than the support itself, while not compromising appreciably its (elasticity modified) Laplace length so that the interactions are long range. The abilities to control the elasticity, the elasto-capillary length and the friction of such a support medium are the unique features of this study that vastly extends those of the generic systems used to study the capillary interactions on liquid surfaces.

We demonstrate the main point of this study by performing a basic experiment in which two equal sized hydrophobized (i.e. PDMS, or polydimethylsiloxane grafted) glass spheres are allowed to interact on the surface of an ultrasoft hydrogel. The field energy of the interaction is estimated from the change in the gravitational potential energy of the particles as they approach each other in conjunction with the well-known superposition principle that was used by Nicolson[7,9] to estimate the same. We validate the underlying scaling analysis with measurements performed with glass spheres of three different diameters, gels of two different moduli as well by reducing significantly the interaction between the particle and the gel by a suitable choice of an external medium. We then show how the knowledge of this field energy of attraction, in conjunction with the speed at which the particles approach each other addresses the role of underlying friction, which is a parameter of importance in soft lubrication. We conclude this study with some demonstrations of how one could begin to study such phenomena as tubulation, phagocytosis and self-assembly in many particles systems.

## 2. RESULTS AND DISCUSSION

### 2.1. Estimation of Energy of Interaction using Gravity

In this research, we use the change of the gravitational potential energy to estimate the energy of interaction of two spheres while they attract each other on the surface of a gel. We begin with the pioneering idea of Nicolson[7] that the driving force behind the interaction of the particles is derived from the distance dependent excess energy of the surface of the gel that results from the deformation induced by two proximous particles. There is a synergistic interplay between this force that is horizontal to the surface and the forces that give rise to the vertical stability of the particles. As the particles tend to sink inside a gel due to gravity and the adhesion forces, the hydrostatic and the shear deformation fields due to surface tension and elasticity resist them. The balance of the above forces dictates the mechanical stability of a particle in contact with gel. Now, as the particles are attracted towards each other, the mechanical fields surrounding them interact, parts of which are cancelled thus resulting in further penetration of the particles inside the gel. We observed a similar phenomenon previously for the particles plunged in a gel[46]. Even when there may be some compensation of the gravitational potential energy due to the modification of the elastic strain energy, as we have shown in a recent paper[46], the overall change of the energy still scales with the former. i.e. $\Delta U \sim -m^* g \Delta h$. This would also be the case with a gel undergoing an elastic deformation due to the combined actions of the gravity and the adhesion forces. To elaborate this point, we consider the case of the small deformation of the gel that can be described by the theory of Johnson, Kendall and Roberts (JKR)[54], where the total energy of the system is:

$$U = \frac{Ka^5}{15R^2} - \frac{P^2}{3Ka} - \frac{Pa^2}{3R} - \pi Wa^2 \quad (1)$$

$$h = \frac{a^2}{3R} + \frac{2}{3}\frac{P}{Ka} \quad (2)$$

Here, $K$ is the contact modulus, $P$ ($\equiv m^*g$) is effective weight of the particle of radius $R$, $a$ is the radius of the contact circle and $h$ is the depth of the penetration of the sphere inside the gel (Figure 4a). Now, using the equilibrium condition: $\partial U/\partial a = 0$, and equation (2) we have:

$$U = -\frac{3m^*gh}{5} - \frac{3\pi}{5}Wa^2 \quad (3)$$

Equation (3) shows that the total energy of the system scales with two quantities: the gravitational potential energy $Ph$ (or $m^*gh$) and the net energy of adhesion $\sim Wa^2$ in the small deformation limit. For a very soft gel such as our system, there is an additional energy term in this equation due to the stretching of the soft solid[55] underneath the particle deforming it. Furthermore, as the gel undergoes a large deformation, the above JKR equation requires modification[56]. Nevertheless, we expect that the total energy will scale as follows:

$$U \sim -m^*gh - \Delta\gamma A_{contact} \quad (4)$$

Where, $A_{contact}$ is the area of contact between the gel and the sphere, and $\Delta\gamma$ is the net change of the interfacial free energy. Furthermore, in the process of the penetration of the particle in the gel, if the contact line is pinned so that the energy due to the stretching of the gel as well as the energy of adhesion remain more or less constant as the particles approach each other, the overall change of the energy of the system would scale simply with the gravitational potential energy $m^*gh$. A variation of $h$ (i.e. $\Delta h$), thus, provides a simple option to gauge how the interaction energy scales with the distance of separation even in the absence of a detailed knowledge of the system. There is also Nicolson's method of expressing the energy of interaction in terms of the separation distance $L$ as $\sim m^* g h_o K_o(L/L_c^*)$, where $L_c^*$ is the elastocapillary decay length that determines the range of interaction and $h_0$ is the vertical distance of the three phase contact line from the undeformed free surface of the gel (figure 4a). The above two forms of estimating interactions energies are equivalent. Thus, to a good order approximation, we may write:

$$m^* g \Delta h \approx m^* g h_o K_o(L/L_c^*) \quad \text{or,} \quad \frac{\Delta h}{h_o} \approx K_o(L/L_c^*) \tag{5}$$

There are certain details about non-linear elasticity that are ignored in the above discussion. We anticipate that an equation of the type described as above would also be valid in a neo-Hookean gel even though the prefactor may be different. Derivation of equation (5) is based upon several approximations and conjectures. We are thus obligated to test its validity experimentally by studying the interaction of two millimeter sized spheres on the surface of low modulus gels where the elastocapillary decay length can be estimated independently. As the particles attract each other, we measure how much they sink ($\Delta h$) inside the gel as a function of $L$. By carrying out experiments with particles of different diameters, using gels of different moduli and virtually switching off the adhesion term by appropriate choice of the surrounding medium, we examine if the data obtained from various experiments would exhibit an universal behavior, i.e. if $\Delta h/h_o$ varies with $L/L_c^*$ following a modified Bessel function of the second kind. In order to achieve the above stated goal, our first objective, however, is to examine how the elastocapillary decay length $L_c^*$ depends on the elasticity of the gel from which to select the appropriate gels for the above stated measurements and analysis.

## 2.2. Capillary Length $L_c$ for water and Elastocapillary Decay Lengths $L_c^*$ for Gels.

Capillary length describes the extent of deformation of a liquid surface due to the competition between the gravity and the surface tension forces. For a liquid surface of surface tension ($\gamma$) and density ($\rho$), this parameter is defined as $L_c = \sqrt{\gamma/\rho g}$, $g$ being the acceleration due to gravity. For pure water, $L_c$ is about 2.7 mm. However, with a hydrogel, this length is expected to be lower than the above value as the elastic modulus of the gel provides additional resistance to surface deformation. We refer this length $L_c^*$ to as 'elasto-capillary decay length' for gels where both elasticity and capillarity contribute. A formal approach to estimate the deformed profile of the surface of the gel would be via a functional minimization of the Lagrangian of the system consisting of the gravitational, surface and the elastic energies with respect to the surface elevation [$\xi(L)$] ( see figure 1a for definition of $\xi(L)$ ), which is postponed for a future publication. We expect that the gravitational and the normal component of the elastic stress would tend to increase the curvature of the gel surface, whereas surface tension and the elastic shear stress would flatten it. The final shape of the surface of the gel is determined by the balance of the above components of stresses, which we examine experimentally and provide a scaling level description of the surface profile as discussed below.

We estimated the elastocapillary decay length by performing experiments with water and gels of different elastic moduli (35Pa to 845Pa), in which a pre-adhered hydrophobic sphere (diameter 2.4mm) was pulled from the surface of either water or a gel and its deformation profile was measured with a microscope. The deformed profile of each of the surface could be fitted with a modified Bessel function of second order of the form $K_0(L/L_c^*)$ from which the characteristic length scale of the deformation $L_c^*$ was obtained (figure 1a). In our hand, the value (2.4 mm) of $L_c$ for pure water is found to be slightly smaller than its theoretical value (2.7 mm). As the surface tension of water used for these measurements is close to the literature value, we ascribe this discrepancy to putative experimental shortcomings. Fortunately, the discrepancy is not significant.

It is reasonable to expect that the value of $L_c^*$ of the elastic gels would be reduced from that ($L_c$) of pure water ($\mu=0$: $\sqrt{\gamma/\rho g}$ ~ 2.4 mm) by a function of the elastocapillary number $\mu\xi_0/\gamma$, i.e. $L_c^* = L_c f(\mu\xi_o/\gamma)$, $\xi_0$ being the maximum elevation of surface (figure 1a). The function should have the following properties: $\mu=0: f(\mu\xi_o/\gamma)=1$, and $\mu \to \infty: f(\mu\xi_o/\gamma) \to 0$. Among various possibilities, a functional form of the type $f(\mu\xi_o/\gamma) \sim \exp[-(\mu\xi_0/\gamma)^n]$, with $n \geq 1$, satisfies the above limits and partially justifies the observation that $L_c^*$ decreases exponentially with modulus, i.e. $L_c^* = L_c \exp(-B\mu)$ (Figure 1b). Figure 1c, however, shows that $L_c^*$ decreases fairly linearly with $\mu\xi_0/\gamma$, i.e.

$$L_c^* \cong L_c\left(1 - \frac{\mu\xi_0}{4\gamma}\right) \qquad (7)$$

Equation 7 does not satisfy the upper limit, i.e. when the shear modulus of the gel becomes extremely large. This equation is likely a low modulus limit of a hitherto undetermined function (figure1c) relating $L_c^*$ and $\mu\xi_0/\gamma$ over a larger range of moduli.

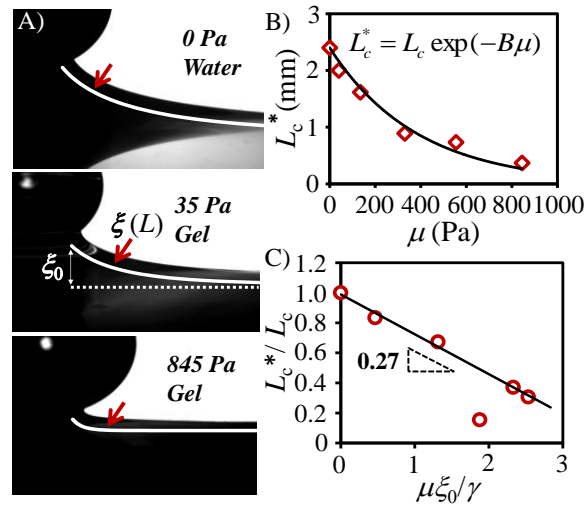

**Figure 1**: (A) Deformed surface profiles of water, and two representative gels are shown. In each case, a 2.4mm diameter PDMS grafted glass sphere was used to deform the surface. The white curves show the fitted Bessel function [ $\xi(L) = \xi_0 K_0(L/L_c^*)$ ] to match the deformed surface profiles. For clarity the fitted curves have been shifted from the deformed surface

profiles along the direction of the red arrows. (B) The plot shows how the elastocapillary decay length decreases with the shear modulus ($\mu$) of the gel. The red open diamonds denote the experimental data and the solid black line is fitted according to $L_c^* = L_c \exp(-B\mu)$ with a value of $B$ as $2.6 \times 10^{-3}$ m$^2$/N. (C) This plot shows that the elasto-capillary decay length decreases with the elasto-capillary number ($\mu\xi_0/\gamma$) for gels of shear modules $\leq 555$ Pa. The datum for a gel of even a higher modulus (845Pa) deviates from the plot.

Note: the underlying section was not part of the original manuscript. After the publication of the paper, we derived a method to estimate Lc* as a function of the modulus of the gel, which is presented below.

---

The deformation of the gel is controlled by the hydrostatic pressure, elasticity of the gel and its surface tension. The normal component of the elastic stress in conjunction with the gravitational pressure tends to curve the gel's surface, whereas the surface tension and the elastic stretching of the gel's surface flatten it. The deformation profile of the gel's surface is thus given by the following equation:

$$\rho g \xi + 2\mu (\partial W / \partial z)_{z=0} = (3\mu H \varepsilon + \gamma)(\partial^2 \xi / \partial x^2) \tag{i}$$

Here, $(\partial W/\partial z)_{z=0}$ is the normal component of the strain at the surface, which is on the order of $\xi\alpha$ ($\alpha = 1/L_c^*$), $H$ is thickness of the active zone of the surface that stretches, which is on the order of $\alpha^{-1}$. The strain $\varepsilon$ of this surface film is proportional to $\xi_o \alpha$. Solution of equation (i) then leads to an expression for $L_c^*$ (or $\alpha^{-1}$) as follows:

$$L_c^* = \frac{(3\mu\xi_o + \gamma)}{\mu + \left(\mu^2 + (3\mu\xi_o + \gamma)\rho g\right)^{0.5}} \tag{ii}$$

A plot showing the variation of $L_c^*$ as a function of the modulus of the film is shown below.

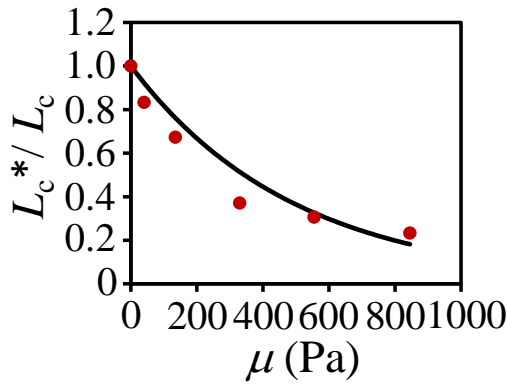

Comparison between the experimental values of $L_c^*$ (solid circles) and those calculated from equation (ii).

---

## 2.3. Attraction of spheres on the surface of gel.

Based on the estimation of the elastic moduli and the elastocapillary decay lengths of gels of different amounts of acrylamide compositions, we found that the gels of modulus ranging from ~ 10 Pa to ~20 Pa are most suitable for studying the elasto-capillary mediated interactions of particles on the surface of a gel. Elastic moduli in this range are significant enough to resist sinking (Figure 3b-d) of the glass spheres of radii ranging from 2 to 4 mm in the gel provided that the glass is made suitably hydrophobic, with which water exhibits a contact angle of ~ 113°. These glass particles, be they hydrophobic or hydrophilic, however, immediately sink in water as their Bond numbers are not significantly lower than unity. The relationship between floatability of a particle and its wettability on a liquid surface has already been discussed previously by Marmur et al.[57], albeit for low Bond number systems. The current observations suggest a worthwhile new direction for these studies when an ultra-low modulus gel is concerned. However, as the focus of our current study is on how the hydrophobic particles interact with each other on the surface of a low modulus gel, the detailed subject of particle floatation by the combined actions of gravity, elasticity and wettability is deferred for a future in-depth study.

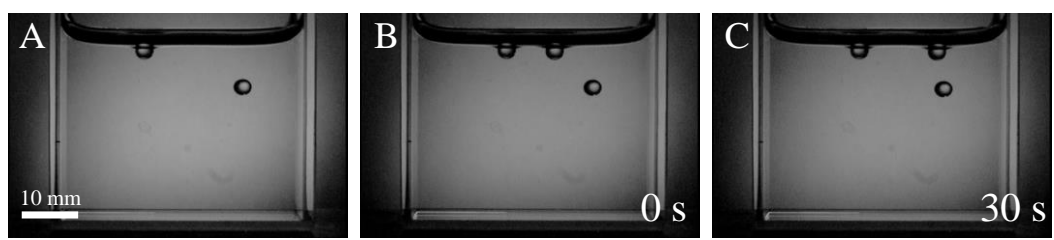

**Figure 2.** (A) Figure illustrating fine balances of the elastic, wetting and gravitational forces giving rise to different types of stabilities of glass spheres (3.2 mm diameter) released on the surface of a polyacrylamide hydrogel of modulus 10 Pa. An untreated (hydrophilic) glass sphere (right) immediately plunges into the gel and becomes neutrally buoyant afterwards. A hydrophobic glass (left) floats on the surface of the same gel. (B) This is an extension of the experiment in A, that shows when another hydrophobic particle is released in between the two, (C) it gets strongly attracted toward the hydrophilic particle, but moves on the gel's surface to minimize its distance of separation from the latter.

The value of $L_c^*$ of a gel of $\mu \sim$ 10 Pa to 20 Pa is large (~2.34mm to 2.28mm) enough that capillary interactions are expected to prevail at a distance comparable to that (2.4mm) of pure water. This hypothesis is tested with a basic experiment, in which we place two glass spheres of equal radii on the surface of a gel at a suitable distance of separation then examine how they attract each other. As stated earlier, a hydrophobic glass particle of any size ranging from 2.4 mm to 4 mm diameter floats on the surface of a PAM gel of modulus ~ 10 Pa to 20 Pa after penetrating the gel partially. When the second sphere is released at a distance of about 1 cm from the first sphere, we studied their mutual attraction on the surface of the gel till they finally come into contact, while penetrating a little further down into the gel yet floating on the surface. Some representative sequences of events of the elastocapillary mediated attraction of particles of three different sizes are shown in figure 3, where the curvature of surface of the gel intervening two spheres are clearly perceptible. Incomplete wrappings of the gel around the PDMS grafted spheres are also evident in these images, a schematic of which is illustrated in figure 4a defining the various parameters that are measured for quantitative analysis of the experimental data. In particular, we measured the average depth of penetration $h$ of both the spheres in the gel with respect to the distance of separation $L$. The depth ($h_0$) of the three phase contact line from the flat surface of the gel was also measured for each experiment and the average value for a set of similar experiments was used for the analysis to follow.

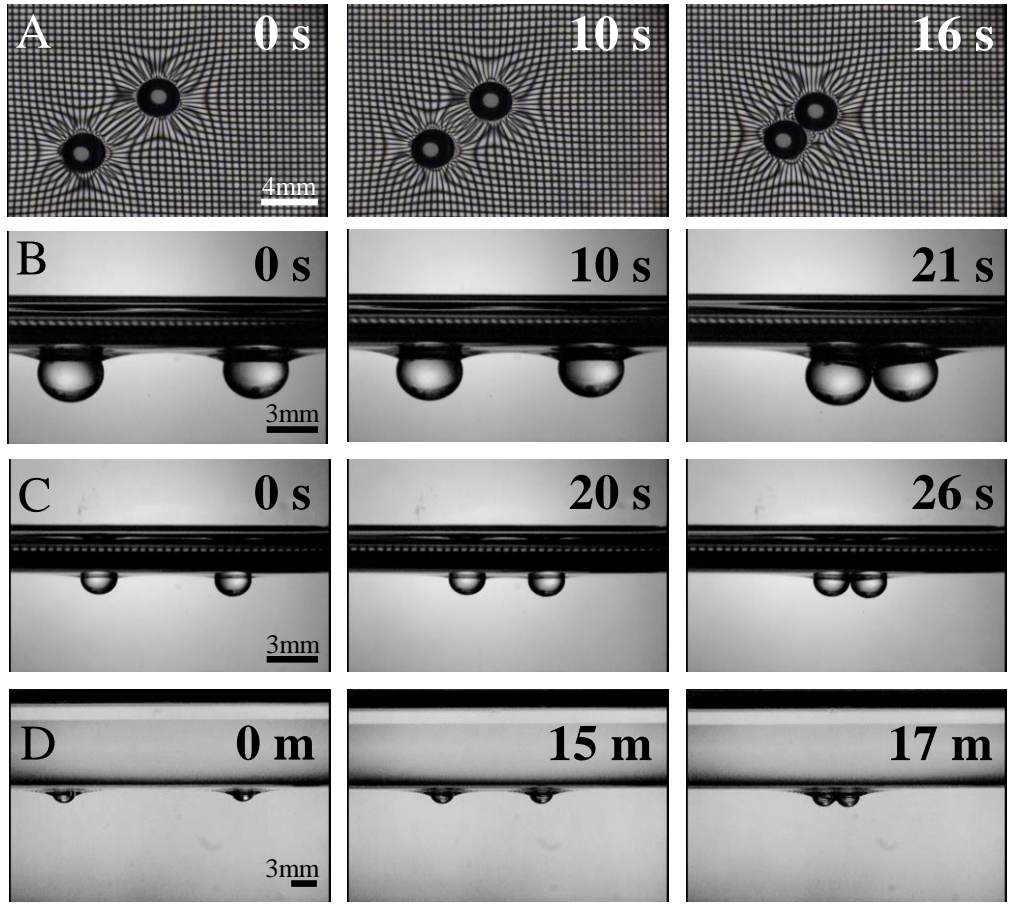

**Figure 3**: (A) A plan view of the interaction of two 3.2 mm hydrophobic glass spheres on the surface of a 10 Pa gel. A wire mesh lined with the base of the glass cell shows the field of deformation of the surface of the gel around the particles. (B) Attraction of the two 4 mm diameter hydrophobic glass spheres on the surface of a 10Pa gel in air (C) Attraction of the two 2.4 mm diameter hydrophobic glass spheres on the surface of a 10Pa gel in air (D) Attraction of the two 2.4 mm diameter hydrophobic glass spheres on the surface of a 10Pa gel in contact with heptane.

For each case, the experiment was repeated at least three times or more to ensure reproducibility. While it is possible to carry out these experiments with a gel of modulus ~140Pa with a significant enough $L_c^*$ (1.62 mm), it was inconvenient to measure $\Delta h$ and $h_o$ accurately as both these parameters strongly decrease with the elastic modulus. We thus restricted our experiments to two gels of moduli 10 Pa ($L_c^* = 2.34$mm) and 19 Pa ($L_c^* = 2.28$mm), the $L_c$ values having been estimated from the empirical relationship obtained in figure 1(b).

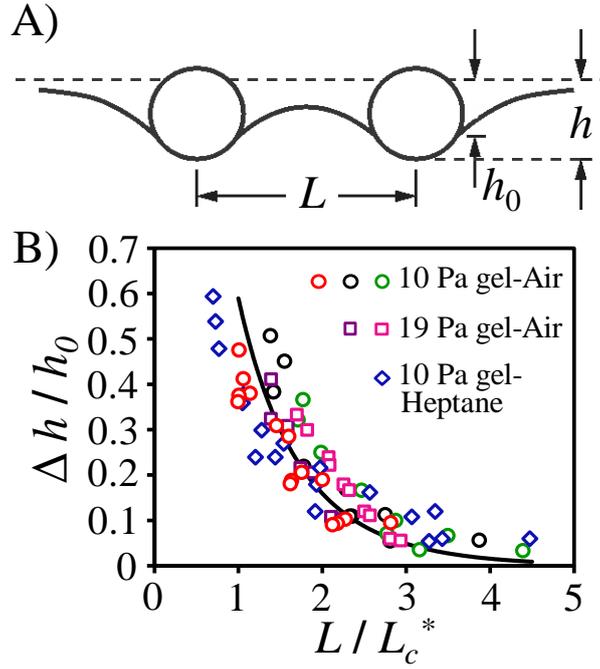

**Figure 4**: (A) Schematic showing elastocapillary attraction of two spheres of identical sizes at a distance $L$. $h$ denotes the depth of submersion of the ball with respect to the initial undeformed level of the gel. The change of $h$ as the spheres approach each other is denoted by $\Delta h$ (see text and figure 4b) $h_0$ is the vertical distance of the three phase contact line from the undeformed free surface of the gel when the spheres are far apart. (B) The change in the depth of separation $\Delta h$ scaled with $h_0$ is plotted as a function of the non-dimensional distance of separation ($L/L_c^*$). Data from all the experiments cluster around the mean curve $\Delta h/h_0 = 1.5 K_o(L/L_c^*)$. The open circles [diameters: 2.4mm (red), 3.2mm (black) and 4mm (green)] denote the 10 Pa gel-Air data. The open squares [diameters: 3.2mm (purple) and 4mm (pink)] denote the 19 Pa gel-Air data. The blue open diamonds (diameter 2.4mm) denote the 10 Pa gel-Heptane data.

In order to compare the results obtained from various experiments, we plotted the normalized average descent of the spheres in the gel, $\Delta h/h_o$, against the normalized separation distance $L/L_c^*$ (figure 4b). It is gratifying that all the data plotted this way cluster around a mean curve $\Delta h/h_0 \propto K_o(L/L_c^*)$ with a proportionality factor of 1.5 that is close to unity expected from a simple model (equation 5) discussed as above.

Note: $h_o$ (figure 4a) has the same physical meaning as $\xi_0$ (figure 1a) in that both describe the vertical distance of the three phase contact line from the undeformed free surface of the gel.

While $\xi_0$ depends on the vertical displacement of the sphere that is controlled externally, $h_o$ is determined by the internal balance of different forces.

**2.4. Attraction of spheres at the interface of n-heptane and gel.**

In all the experiments described in section 2.3, adhesion plays a significant role in the sense that the gel wraps around the particle so much that the effective $h_o$ is rather small. One can surmise that if the interfacial adhesion is decreased, the gel would undergo mainly a Hertzian deformation meaning that the sphere mildly deforms the gel so that the value of $h_o$ would increase. The adhesion energy in this system could indeed be reduced considerably by replacing the upper air-layer over gel with n-heptane. The free energy of adhesion between a PDMS grafted glass particle with the gel is[58]:

$$\Delta G_{132} = -2(\sqrt{\gamma_1} - \sqrt{\gamma_3})(\sqrt{\gamma_2^d} - \sqrt{\gamma_3}) \qquad (8)$$

Where, $\gamma_1$ is the surface energy (totally dispersive) of PDMS (~ 22 mN/m), $\gamma_2$ is the dispersion component of the surface energy (~ 21.8 mN/m) of the gel, and $\gamma_3$ (20.1 mN/m) is the surface tension (totally dispersive) of n-heptane. The adhesion energy of the hydrophobic glass particle with the gel is thus estimated to be - 0.08 mJ/m$^2$ which is negligible compared to the expected value of -44mJ/m$^2$ in air. $h_0$ (0.87 mm) of a small particle released on the surface of the gel of modulus 10 Pa through heptane is indeed found to be larger than that (0.32 mm) in air. $L_c^*$ of the gel-heptane interface was measured by releasing a 4 mm diameter hydrophobic glass sphere at the interface and measuring the deformed profile of the same as it bent towards the gel. This contrasts the method used in air, where a pre-adhered sphere was pulled away from the gel-air interface (Figure 1a). The deformed profile at the gel-heptane interface could also be fitted with a modified Bessel function thus yielding an elastocapillary decay length of 3.5 mm, which is found to be larger than the value (2.34mm) of the same gel in contact with air. It is reassuring that the above value of $L_c^*$ is close to that (3.9 mm) estimated using the empirical relationship $L_c^* = L_c \exp(-B\mu)$ (figure 1b), in which $L_c = \sqrt{\gamma_{int}/\Delta\rho g}$ is calculated using an interfacial energy ($\gamma_{int}$) of the gel-heptane interface as 51mN/m (~ the value at heptane-water interface) and $\Delta\rho$ (the difference in the densities of water and heptane) as 316 kg/m$^3$. These measurements exemplify that the elastocapillary decay length of an interface can indeed be modified by its surface tension. As a consequence of both larger values of $h_o$ and $L_c^*$, the spheres recognize each other at a distance much larger than they do on the gel in contact with air. The strength of the interaction is also larger as evidenced from the fact that the net change of the gravitational potential energy for the spheres in going from an infinite separation distance to contact at heptane-gel interface is three times larger than that at air-gel interface. However, the normalized descent of the spheres in the gel, $\Delta h/h_o$, when plotted against $L/L_c^*$ (figure 4b) still cluster around the same mean curve obtained from the experiments at air-gel interface.

## 2.5. Role of friction in elasto-capillary attraction: Difference at Gel-Air and the Gel-heptane interfaces

Apart from quantifying how the surface tension and the elastic forces play joint roles in determining the energy of interaction of two spheres on the surface of a soft gel, these experiments also have fascinating prospects in studying how the coupled elastic and hydrodynamics forces play their roles in determining the friction of the spheres with gel. This subject is of interest in the field of so called[52] "soft lubrication", in which both elasticity and hydrodynamics play their respective roles. We begin with the asymptotic logarithmic form of the energy of interaction of the spheres as given in equation (5), the derivative of which gives the force of interaction as: $m^*gh_0/L$. If we assume that this force is balanced by a kinematic resistance that is linearly proportional to velocity, we have:

$$\zeta(-dL/dt) \sim m^*gh_0/L \qquad (9)$$

$$\text{Or,} \quad -L^2/t \sim m^*gh_0/\zeta \qquad (10)$$

Where, $\zeta$ is the kinematic friction coefficient. Equation 10, in which the negative effective diffusivity ($L^2/t$) is inversely proportional to friction coefficient, pleasantly is the deterministic analog of the well-known Einstein-Sutherland's equation[59] connecting diffusivity and friction. Here, $m^*gh_0$ is equivalent to an "effective gravitational temperature" introduced earlier by Segre et al[60] in the context of sedimentation of particles. According to equation 10, a plot of $L^2$ versus $t$ should be linear, the slope of which is inversely proportional to the kinematic friction coefficient. The dynamics of the attraction of spheres at the gel-heptane interface is quite different from that at the air-gel interface in that the time taken for the two spheres to come into contact in the former case is much larger than that at the gel-air interface. $L^2$ is fairly linear with $t$ in both cases (figure 5), which further ascertains that the friction is fairly linear with velocity and that the form of the attractive field energy is asymptotically logarithmic at a short separation distance. The slope of the $L^2$-$t$ for the heptane-gel interface (0.25 mm$^2$/s) is, however, an order of magnitude smaller than that at the air-gel interface (2.3 mm$^2$/s) thus ascertaining the large differences in the friction in both cases. With the appropriate values of $h_o$ and the fair estimations of the buoyancy corrected weights of the spheres in the two cases, we estimate that the kinematic friction coefficient at the heptane-gel interface is about 20 times larger than that at the gel-air interface. The absolute value of $\zeta$ for the 2.4 mm diameter spheres attracting at the heptane-gel interface (equation 10) is estimated as 0.7 Nm$^{-1}$s$^{-1}$ whereas that from Stokes drag force ($\sim 6\pi\eta R$) is estimated as 8.7x10⁻⁶ Nm⁻¹s⁻¹ in Heptane. This large discrepancy of the values of $\zeta$ suggest a complex origin of friction, part of which could be arising from the deformation of the gel and part due to the wedge flow of the liquid heptane near the sphere-gel interface. The friction coefficient for the spheres undergoing attraction at the gel-air interface is estimated to be 0.03 Nm⁻¹s⁻¹. In our previous study involving the attraction of particles that were completely submerged inside the gel, the friction coefficient was estimated to be 1.7 Nm⁻¹s⁻¹,[46] which is much higher than the friction coefficient value observed for the spheres in the present study where they are only partially wrapped with the gel surface.

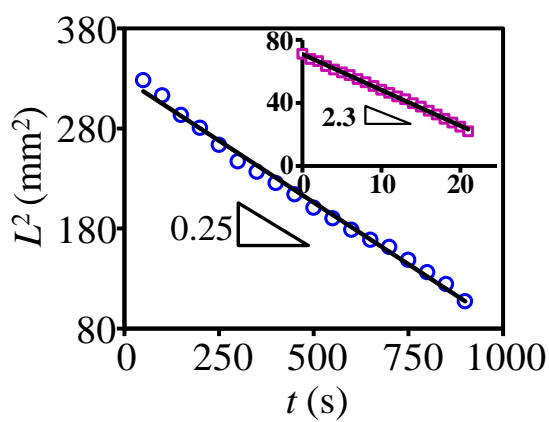

**Figure 5:** This plot shows that the squared distance of separation between two interacting particles decreases linearly with time. The blue open circles correspond to the 2.4mm diameter spheres at the gel- heptane interface. (Inset) The pink open squares correspond to the 2.4mm diameter spheres at the gel- air interface. The shear modulus of the gel in each case is 10 Pa.

**2.6. Tubulation and Self-Assembly of spheres at the n-Heptane and Gel interface**

The long range interaction of the particles in a gel coupled with the fine balance of forces that give rise to their stability perpendicular to the surface of the gel, can give rise to several interesting scenarios mimicking phenomena in biological and other settings[50] that are worth studying in detail. Here we provide two examples: tubulation and self-assembly of the particles at the heptane-gel interface.  In our experiments, tubulation has frequently been observed with a very soft gel (10Pa) when covered with a layer of n-heptane that reduces the adhesion energy and thus the propensity for the gel to completely wrap around the particles.

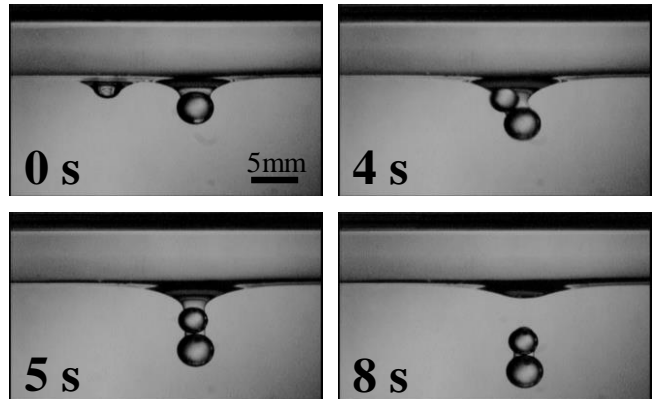

**Figure 6**: A 3.2mm diameter hydrophobic glass sphere is attracted towards a 4mm size hydrophobic glass sphere. As the two spheres contact each other, the pair re-orients inside the tube adjoining the two (tubulation). Finally, the pair penetrates inside the gel and becomes stagnant to a point where it becomes elasto-buoyant.

When a large particle (4 mm) is first introduced on the surface of such a gel, it attains its stability after penetrating the gel to a significant distance, whilst still remaining on its surface (Figure 6). However, when a smaller particle (3.2 mm) is released at a distance of about 9 mm from this particle, it traverses on the surface to reach the larger particle due to elastocapillary attraction. As the two particles touch each other, the pair re-orients and inserts itself in the gel till the accumulated strain induced elastic force balances the combined weight

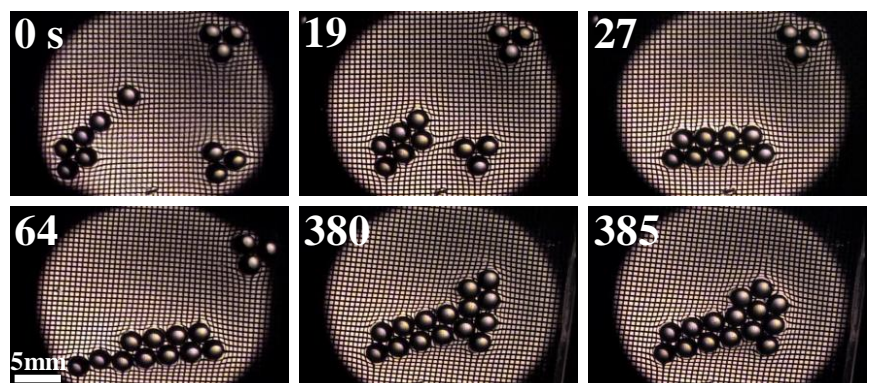

**Figure 7**: A plan view of elastocapillary surface force mediated self-assembly of 2.4mm silanized glass spheres at the interface of a gel (10Pa) and n-heptane. As the spheres are randomly dispersed on the surface of the gel through heptane, they form random clusters,

which then move towards each other forming one large cluster. A wire mesh lined with the base of the glass cell shows the field of deformation of the gel surface.

of the particles. The sequence of events mimics what is known as tubulation in biomembrane system, which underlies the interactions and penetrations of nanoparticles through a biological membrane[49]. However, it is important to emphasize that the range, the strength and the origins of interactions here are quite different from what is observed in typical biomembrane sytems. What is presented here is only a mimicry of the biological phenomena in a mesoscale level system.

Another observation is the self-assembly in a many particle system that is observed when several small particles are dispersed on the surface of the gel under heptane (figure 7). The experiments shown in the pictures were performed using silanized glass spheres, although same kind of interfacial self-assembly was also observed with PDMS grafted glass spheres. The particles attract and form various structures resembling chains and triangles. These clusters eventually are attracted towards each other thus forming larger aggregates. These types of phenomena are generic with particles dispersed on the surface of liquid, such as water. However, here, the ability to control the range and the strength of the interaction through the modulation of elasticity, the interaction forces, as well as the friction, the parameter space of investigation can be vastly enhanced. Thereby, these systems could potentially be used to study nucleation and clustering phenomena in a 2d system more rigorously than what might be possible with a liquid surface alone.

## 3. REITERATING MAIN POINTS

The elastocapillary force mediated attraction of particles on the surface of an ultra-soft gel follows a modified Bessel function of second kind (zeroth order) much like the generic capillary attractions of particles on the surfaces of liquids. For an elastic hydrogel, however, the range of interaction is reduced by its elasticity, which has implications in designing systems where the range of attraction can be tuned in by the elasticity of the support. A simple relationship between the descents of the particles in the gel with their distance of separation was developed by equating the change of the gravitational potential energy of the attracting spheres with a well-known form of interaction obtained from the superposition principle of Nicolson. Though shrouded by some uncertainties of the implicit assumptions and approximations, the proposed relationship stood firmly against the experimental tests involving spheres of different sizes suspended on gels of different moduli, even when the adhesion of a gel/particle interface is almost non-existent. This experimental assertion gives us confidence to consider that the main physics underlying the interactions of particles on the surface of an elastic gel has been essentially captured. There are, nevertheless, some differences in the data obtained from one set of experiment to another. While, part of this discrepancy is due to experimental uncertainties, a more precise analysis of the data by taking into consideration the second order effects resulting from the stretching of the gel, its large deformation neo-Hookean behavior and some change of the interfacial energy in those cases where the contact lines are not entirely pinned on the descending spheres may also be required.

In all cases, the particles approach each other with a negative effective diffusivity that is a deterministic analog of the Einstein-Sutherland's relationship. This is a consequence of an attractive force varying inversely linearly with the separation distance and the dynamics of motion being governed by a linear kinematic friction. A surprising observation of this study is that the kinematic friction of the sphere on the gel in contact with heptane is substantially larger than that in air. This, we believe, presents itself as an important problem of soft lubrication that deserves detailed in-depth investigation. While in a thermal system, the particles would move away from each other, here they come closer, i.e. they diffuse negatively with an effective gravitational temperature. An effective gravitational temperature was introduced earlier[60] in the context of the sedimentation of the particles. However, as the "effective gravitational temperature" as introduced here does not possess the feature of fluctuation that allows particles to explore the entire phase space, no major issue is to be made of out of it other than treating this quantity as an "intensive property" of the system that gives rise to an effective diffusion like kinetics. We expect that the putative analogy may be more useful for an ensemble of a large numbers of particles, where they could form random clusters and could even move against a concentration gradient. This particular feature could be interesting in setting up experiments to study clustering phenomena and phase separation kinetics in mixed particle systems.

Elucidation of the nature of friction in such systems is an important challenge, a satisfactory resolution of which would require full knowledge of whether the spheres rotate or slip, and if the wedge flow of the surrounding liquid through the gap formed by the sphere on the gel (that has a diverging stress) plays any role. Preliminary studies with the attracting spheres do not show any clear evidence of rotation as they approach each other, even though, surprisingly, some azimuthal motion of the spheres was noticeable. The subject promises rich underlying physics of soft lubrication, which is a coupled problem of hydrodynamic flow and elastic deformation. Nevertheless, the fact that friction can be modified in such systems provides another avenue to manipulate the dynamics of the motion of the particles as much as the range and the strength of interaction can be modulated by appropriate choices of the surface tension and the elasticity of the gel itself.

We end this section by commenting that what we learned here involving the interaction of particles on the surface of a gel can be combined with what we reported earlier[46], namely the interaction of particles deeply plunged inside a gel in order to enhance the overall scope of the elasto-capillary mediated interactions of particles in a gel. This philosophy can be illustrated with a simple example described in figure 2, in which a hydrophilic glass particle plunges inside a gel and a hydrophobic particle released far away from this one floats on the gel's surface. If however, a second hydrophobic particle is released in between the two, it gets attracted more strongly towards the hydrophilic particle, but moves on the gel's surface in order to minimize its distance from it. This observation, in which a floating particle interacts with the strain field produced by a submerged particle, may give rise to new twists to particle interactions that have not been exploited thus far. Thus, artificially created elasto-capillary field inside a gel may be a novel way to assemble particles on its surface that may even be extended to those of colloidal dimensions.

4. SUMMARY AND CONCLUSION.

This study shows that elastic forces coupled with surface tension are advantageous in studying the attraction mediated 2d organization of particles on a soft support. Dense particles can be easily dispersed on such a support that would otherwise sink in normal liquids. Further control of the particle interaction, i.e. its range, strength and dynamics, can be easily achieved. It should also be possible to support a relatively thin (~ few millimeter) layer of hydrogel on an elastomeric support, e.g. a crosslinked polydimethyl siloxane, that can be stretched or compressed uniaxially or biaxially to induce surface folding, or in which patterns can be formed via an external field, in order to add additional degrees freedom to manipulate interactions and self-assembly of particles suspended on the surface of the gel. This configuration could be particularly useful in studying the elasticity and the buckling transitions of particle rafts formed by self-assembly. These studies are poised for further explorations in many particle systems to gain deeper understanding of such phenomena as clustering, jamming, tubulation and possibly phase separation kinetics in mixed systems.

5. EXPERIMENTAL DETAILS

Although the procedure for the preparation of the gel has been described in our recent publications[45,46], we provide that information with some additional details in the supporting information (SI) section. Here, we discuss the essential details related to the current studies.

### 5.1. Measurement of Shear Moduli of the Polyacrylamide Gel.

The shear moduli of the gels were estimated with a slight modification of a previously reported method [45], in which the gel was forced to undergo a shear vibration by means of a random external field (figure 8). Please refer to SI for rest of the details.

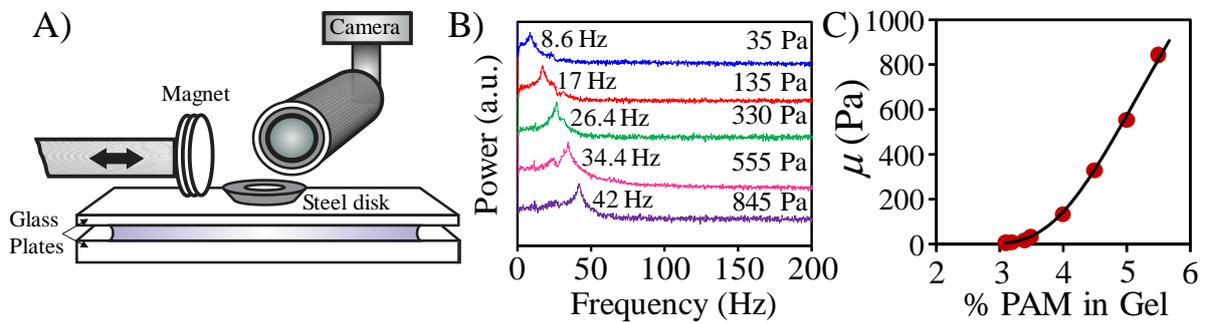

**Figure 8**: (A) Schematic of the method used to measure the shear modulus of a gel. The method involves the creation of a random magnetic field that interacts with the steel disk and vibrates the upper glass plate randomly with respect to the lower plate thus creating a random shear deformation of the gel. (B)The resonance peak of the shear vibration of the gel was obtained from the power spectra of its random vibration with which the shear moduli were calculated using $2\pi\omega = \sqrt{\mu A / mH}$ . (C) The shear moduli ($\mu$) of different gels plotted as a function of the percentage of polyacrylamide ($x$) in them follows an empirical relationship $\mu = 2500\exp(-178/x^{2.98})$.

### 5.2. Measurements of $L_c$ (Laplace Length) for water and $L_c^*$ (Elasticity modified Laplace Length) for Gels.

The elastocapillary decay length $L_c^*$ of water or a gel was estimated as follows. For water, a clean polystyrene petri dish (100mm $\phi$) was placed on a stage in front of a CCD (charged-coupled device) camera (RU, Model: XC-75) atop a vibration isolation table and was filled with DI water till the brim. A hydrophobic glass sphere (2.4mm $\phi$) was attached on the bottom of a stage that could be moved up and down using a micromanipulator. The sphere was brought down very close to the surface of the water until its reflection was visible on the surface of the water. By knowing the distance between the object and the image, we could ascertain rather precisely the reference level of the water from the midpoint of the object and its image. The camera was focused in such a way that the sphere was on one side of the frame such that a large area of the deformed surface could be captured. The glass sphere was then brought downward till it touched the surface of the water resulting in a concave meniscus. The sphere was then slowly moved upward while the whole process was captured in the form of a video by using the CCD camera and recorded with WinTV on the computer. The video was decomposed into image sequence in VirtualDub. All the image analysis were performed using ImageJ. The deformed profile was plotted as $\xi(L)$ versus $L$, where $L$ is the distance measured from the vertical line passing through the center of the glass sphere in the image and $\xi(L)$ is the vertical distance of the deformed profile measured from the undeformed reference level (figure 1a). The profile of the deformed surface of water could be fitted with a modified Bessel function $K_0(L/L_c)$ of the second kind where $L_c$ is the capillary or Laplace length of water (figure 1a). The same process of measuring the $L_c^*$ was repeated with physically cross-linked polyacrylamide gels of concentrations 3.5%, 4%, 4.5%, 5% and 5.5% of the acrylamide monomer having the shear moduli ranging from 35 Pa to 845 Pa. The gels were cured in clean polystyrene petri dishes and kept inside a large glass petri dish with stacks of DI water-soaked filter papers for 24 hours. The round gel slab was overturned into the lid of the petri dish so that we could measure the $L_c^*$ of the gel's reverse flattened side (Figure 1b,c). We repeated the measurements of $L_c^*$ on the top surface of the cured gel slab in the petri dish that also gave almost similar values with the experiments done with the reverse side that is well within the experimental error.

**5.3. Attraction of spheres on surfaces of gels.**

In order to study the attraction of hydrophobic glass spheres on the surface of gel, we placed the home built glass cell containing the cured gel on a stage in front of a CCD camera (MTI, CCD-72). All the experiments were performed followed by 24hrs of the curing the gel, which ensured evaporation of extraneous water from its surface, although we suspect that a very thin layer of water still remains on the surface of the gel. All the experiments were performed on the flattest parts of the gels to eliminate any putative artefacts arising from the curvature of the gel meniscus close to the walls of the test cell. A hydrophobic glass sphere was first released on the gel surface followed by the release of a similar sized sphere within ~10mm away from the first one. Their attraction was recorded using a CCD camera that was attached to a variable focal length microscope (Infinity). Similar experiments were also performed at the interface of gel and n-heptane. When the spheres were released on the surface of n-heptane, they sunk through heptane and rested at the interface formed between n-heptane and the gel. The attractions were recorded in the CCD camera and analyzed using ImageJ. All the experiments were repeated at least three or more times. A stainless steel (SS316) wire cloth (opening size 0.015", wire diameter 0.010") was lined with the base of the glass cell to observe the deformations in the gel as the particles interacted. At this point, we have not analyzed the optical distortions of the wire mesh; however they can be analyzed with a ray tracing software to quantify the deformations of the surface of the gel (figures 3a and 7).